\documentclass[prl,twocolumn,showpacs,superscriptaddress]{revtex4}
\usepackage{graphicx}
\usepackage{indentfirst}
\usepackage{epsfig}
\usepackage{subfigure}
\usepackage{amsmath}
\usepackage{amssymb}
\def\be{\begin{eqnarray}}
\def\ee{\end{eqnarray}}

\begin{document}
\title{Dipole emission and coherent transport in random media II.\\
    Density of states vs. index of refraction}
\author{M. Donaire}
\email{manuel.donaire@uam.es}
\address{Departamento de F\'{\i}sica de la Materia Condensada, Universidad Aut\'{o}noma de
Madrid, E-28049 Madrid, Spain.}
\begin{abstract}
This is the second of a series of papers devoted to develop a
microscopical approach to the  dipole emission process and its
relation to coherent transport in random media. In this Letter, we
deduce a relation between the transverse decay rate of an emitter
in a virtual cavity and the complex refraction index of the host
medium. We argue on the possibility of a criterion for
inhibition/enhancement of spontaneous emission in function of the
transition frequency and the correlation length of the host
scatterers. In addition, we study the radiative/non-radiative
nature of the net power emission through a microscopical analysis
of the scattering events involved. This study reveals essential
discrepancies with previous interpretations.
\end{abstract}
%
 \maketitle
\indent In a previous work \cite{paperI} we computed the average
local density of states (LDOS) within an isotropic and homogeneous
medium. It is the aim of this Letter to use those equations to
deduce a relation between LDOS and coherent transport parameters.
Recent experiments \cite{Wiersma} have put forward evidences of
such a relation. In \cite{paperI}, several scenarios were
addressed for an emitter seated in either a virtual or a real
cavity. In this Letter we focus on a virtual cavity scenario. We
will consider the host medium made of a set of spatially
disconnected point scatterers of density $\rho$ and the emitter as
a fluorescence particle seated on top of a host scatterer. We will
refer to that scatterer as \emph{host particle}. The emitter
transition frequency and the resonance frequency of the host
particle are assumed far apart from each other. Dipole induction
is completely cast in the host particle polarizabiliy. A bulk
propagator $\bar{G}$ and electrical susceptibility $\bar{\chi}$
can be unambiguously defined, where overlines denote rank-2
tensors. The decay rate $\Gamma$ reads \cite{paperI}
\begin{eqnarray}
\Gamma^{\alpha,\mu}_{\omega}&=&\frac{-2\omega^{2}}{3\hbar\epsilon_{0}c^{2}}\frac{|\mu|^{2}}
{|1+\frac{1}{3}k_{0}^{2}\alpha_{0}[2\gamma_{\perp}+\gamma_{\parallel}]|^{2}}
\Bigl[\Im{\{2\gamma_{\perp}+\gamma_{\parallel}\}}\label{latercera1}\nonumber\\
&-&\frac{k_{0}^{2}}{3}\Im{\{\alpha_{0}\}}|2\gamma_{\perp}+\gamma_{\parallel}|^{2}\Bigr]\label{latercera2},
\end{eqnarray}
where the factors $\gamma_{\perp,\parallel}$ are given by
\begin{eqnarray}
2\gamma_{\perp}&=&
\int\frac{\textrm{d}^{3}k}{(2\pi)^{3}}\Bigl[\frac{2\chi_{\perp}(k)/(\rho\alpha)}{k_{0}^{2}[1+\chi_{\perp}(k)]-k^{2}}\Bigr]-2\Re{\{\gamma^{(0)}_{\perp}\}},
\label{LDOSIper}\\
\gamma_{\parallel}&=&\int\frac{\textrm{d}^{3}k}{(2\pi)^{3}}\:\Bigl[\frac{1}{\rho\alpha}\frac{\chi_{\parallel}(k)}{k_{0}^{2}[1+\chi_{\parallel}(k)]}-\frac{1}{k_{0}^{2}}\Bigr].\label{LDOSIparal}
\end{eqnarray}
In the above formulae, $\alpha_{0}\equiv 4\pi a^{3}\frac{\epsilon_{e}-1}{\epsilon_{e}+2}$ is the electrostatic polarizability of the host scatterers,
being $\epsilon_{e}(\omega)$ the single-particle dielectric contrast; $\alpha=\frac{\alpha_{0}}{1-\frac{i k_{0}^{3}}{6\pi}\alpha_{0}}$ is the in-vacuum
 renormalized single-particle polarizability; $\mu$ is the spontaneous transition amplitude of the emitter and
$\omega$ the transition frequency, with $k_{0}=\omega/c$. The term
$-2\Re{\{\gamma^{(0)}_{\perp}\}}$ accounts for the singularity of
the radiative part of $\bar{G}^{(0)}(r)$ at $r=0$ in vacuum
(denoted by the script (0)). The denominator in
Eq.(\ref{latercera2}) amounts to self-polarization cycles over the
host particle. It gives rise to an effective renormalization of
$|\mu|^{2}$, $|\tilde{\mu}|^{2}\equiv \frac{|\mu|^{2}}
{|1+\frac{1}{3}k_{0}^{2}\alpha_{0}[2\gamma_{\perp}+\gamma_{\parallel}]|^{2}}$,
which depends weakly on $\gamma$ factors. In a first approach, we
can distinguish three contributions to
$\Gamma^{\alpha,\mu}_{\omega}$ --we drop the scripts on $\Gamma$
hereafter, namely
\begin{eqnarray}
\Gamma^{0}_{abs.}&=&\frac{2\omega^{4}}{9\hbar\epsilon_{0}c^{4}}|\widetilde{\mu}|^{2}|2\gamma_{\perp}+\gamma_{\parallel}|^{2}\Im{\{\alpha_{0}\}},\label{la1}\\
2\Gamma_{\perp}&=&\frac{-4\omega^{2}}{3\hbar\epsilon_{0}c^{2}}|\widetilde{\mu}|^{2}
\Im{\Bigl\{\int\frac{\textrm{d}^{3}k}{(2\pi)^{3}}\:
\frac{\frac{\chi_{\perp}(k)}{\rho\alpha}}{k_{0}^{2}[1+\chi_{\perp}]-k^{2}}\Bigr\}},\label{la2}\\
\Gamma_{\parallel}&=&\frac{-2\omega^{2}}{3\hbar\epsilon_{0}c^{2}}|\widetilde{\mu}|^{2}
\Im{\Bigl\{\int\frac{\textrm{d}^{3}k}{(2\pi)^{3}}\:
\frac{\frac{\chi_{\parallel}(k)}{\rho\alpha}}{k_{0}^{2}[1+\chi_{\parallel}]}\Bigr\}}.\label{la3}
\end{eqnarray}
Eq.(\ref{la1}) is the decay rate due to absorbtion within the host particle. The
other two equations correspond to transverse and longitudinal
decay respectively. Propagating modes are determined by the
singularities of the integrands in Eqs.(\ref{la2},\ref{la3}) in
the complex plane, that is, poles and branch cuts. Poles are modes
$k_{\perp},k_{\parallel}$ satisfying the dispersion relations in
the effective medium,
\begin{equation}\label{dispersion}
k^{2}-k_{0}^{2}[1+\chi_{\perp}(k)]|_{k=k_{\perp}}=0,\qquad
1+\chi_{\parallel}(k)|_{k=k_{\parallel}}=0.
\end{equation}
The refraction index, $\bar{n}$, and the extinction mean free path, $l_{ext}$, parametrize propagation over long distances. That is, they correspond to the longest wave length solutions for $k_{\perp}$.
 We will disregard the possibility of propagating longitudinal modes as
these do not affect directly the relation between $\Gamma$ and the
index of refraction. As a matter of fact, propagation of
longitudinal modes requires some fine tuning on the arrangement of
scatterers \cite{Citrin}. Because of isotropy, $\bar{\chi}$ is a
function of $k^{2}$. It is only for $k_{0}\xi\ll1$ that this
dependence is weak, $\xi$ being the typical correlation length
between host scatterers. For a virtual cavity, the cavity radius
$R$ can be identified with $\xi$ as well.  It is for
$k_{0}\xi\ll1$ that $\bar{\chi}$ admits an expansion in powers of
$k^{2}/k_{0}^{2}$ and it suffices to truncate the series at order
$k^{2}/k_{0}^{2}$. In doing so, Eq.(\ref{la2}) presents a unique
pole $k_{\perp}$ and the whole transverse propagator is
characterized solely by $\bar{n}$ and $l_{ext}$,
$\bar{n}=\Re{\{\sqrt{1+\chi_{\perp}(k_{\perp})}\}}$,
 $l_{ext}^{-1}=2k_{0}\Im{\{\sqrt{1+\chi_{\perp}(k_{\perp})}\}}$.
On the other hand, we can distinguish two components in
$\Gamma_{\perp}$, namely,
\begin{eqnarray}
2\Gamma_{\perp}^{NP}=\frac{-4\omega^{2}|\widetilde{\mu}|^{2}}{3\hbar\epsilon_{0}c^{2}}
\int\frac{\textrm{d}^{3}k}{(2\pi)^{3}}
\Im{\{\frac{\chi_{\perp}}{\rho\alpha}\}}\Re{\Bigl\{\frac{1}{k_{0}^{2}[1+\chi_{\perp}]-k^{2}}\Bigr\}},\label{lb1}\\
2\Gamma_{\perp}^{P}=\frac{-4\omega^{2}|\widetilde{\mu}|^{2}}{3\hbar\epsilon_{0}c^{2}}
\int\frac{\textrm{d}^{3}k}{(2\pi)^{3}}
\Re{\{\frac{\chi_{\perp}}{\rho\alpha}\}}\Im{\Bigl\{\frac{1}{k_{0}^{2}[1+\chi_{\perp}]-k^{2}}\Bigr\}},\label{lb2}
\end{eqnarray}
where the scripts $P$ and $NP$ stand for \emph{propagating} and
\emph{non-propagating} respectively. In Eq.(\ref{lb2})  the
'normalized' susceptibility function
$\Re{\{\frac{\chi_{\perp}(k)}{\rho\alpha}\}}$ plays a role
analogous to that of the field-strength renormalization factor $Z$
in a quantum field theory (QFT). In QFT  it is the analytical
structure of  $G(k)$ together with $Z$ in the complex plane that
determines the mass spectrum of quantum particles propagating in
space-time \cite{Peskin}. Correspondingly, it is the analytical
structure of $G_{\perp}(k)$ together with
$\Re{\{\frac{\chi_{\perp}(k)}{\rho\alpha}\}}$ that determines the
spectrum of propagating coherent photons in a random medium. In
view of the above classification, we will refer to
$\Gamma_{abs.}^{0}$, $\Gamma_{\parallel}$ and
$2\Gamma_{\perp}^{NP}$ all together as \emph{non-propagating decay
rate},
$\Gamma^{NP}=\Gamma_{abs}^{0}+\Gamma_{\parallel}+2\Gamma_{\perp}^{NP}$;
while $2\Gamma_{\perp}^{P}$ is the only propagating component,
$\Gamma^{P}=2\Gamma_{\perp}^{P}$ --compare with other
interpretations \cite{LagvanTig}. It is however not clear, except
for $\Gamma_{abs.}^{0}$, which contribution of $\Gamma^{NP}$ is
associated to absorbtion by host scatterers and which part
corresponds to incoherent radiation. That needs a deeper
microscopical analysis. We proceed to compute $\Gamma_{\perp}^{P}$
in function of the extinction coefficient,
$\kappa=l^{-1}_{ext}/(2k_{0})$, and $\bar{n}$. On the one hand, we
have that
$\Re{\{\chi_{\perp}(k_{\perp})\}}=\bar{n}^{2}-1-\kappa^{2}$ and
$\Im{\{\chi_{\perp}(k_{\perp})\}}=2\bar{n}\kappa$. On the other
hand, Eq.(\ref{lb2}) yields
$\frac{\omega^{3}|\widetilde{\mu}|^{2}}{3\pi\hbar\epsilon_{0}c^{3}}\Re{\{\frac{\chi_{\perp}(k)}{\rho\alpha}\}}\:\bar{n}$.
Putting everything together we end up with
\begin{eqnarray}\label{theone}
2\Gamma_{\perp}^{P}&=&\frac{2\omega^{3}|\widetilde{\mu}|^{2}}{3\pi\hbar\epsilon_{0}c^{3}}\:\bar{n}\:\Re{\{\chi_{\perp}/\rho\alpha\}}\nonumber\\
&=&\frac{2\omega^{3}|\widetilde{\mu}|^{2}}{3\pi\hbar\epsilon_{0}c^{3}\rho|\alpha|^{2}}\:\bar{n}\:\Bigl[
\Bigl(\bar{n}^{2}-1-\kappa^{2}\Bigr)\Re{\{\alpha\}}\nonumber\\&+&2\bar{n}\kappa\Im{\{\alpha\}}\Bigr].
\end{eqnarray}
Taking the approximation $\epsilon\approx \bar{n}^{2}$ and
assuming that Maxwell-Garnett  relation (MG) holds,
$\epsilon=1+\frac{\rho\alpha}{1-\frac{1}{3}\rho\alpha}$ ,
Eq.(\ref{theone}) goes as
$\sim\frac{\epsilon+2}{3}\sqrt{\epsilon}$. The missing local
factor with respect to the usual virtual cavity formula
\cite{Lorentz,LoudonI,Scheel} implies that only part of the decay
rate there is transverse and radiative. We will go over this point
later on. Proceeding in a similar manner, $2\Gamma_{\perp}^{NP}$
reads
\begin{eqnarray}
2\Gamma_{\perp}^{NP}&=&\frac{-2\omega^{3}|\widetilde{\mu}|^{2}}{3\pi\hbar\epsilon_{0}c^{3}}\Bigl[\kappa\Im{\{\chi_{\perp}/\rho\alpha\}}
-\frac{\bar{n}\kappa}{2k_{0}\xi}\Bigr]\label{lalo}\\
&=&\frac{2\omega^{3}|\widetilde{\mu}|^{2}}{3\pi\hbar\epsilon_{0}\rho|\alpha|^{2}c^{3}}\kappa\Bigl[
\Bigl(\bar{n}^{2}-1-\kappa^{2}\Bigr)\Im{\{\alpha\}}\nonumber\\&-&2\bar{n}\kappa\Re{\{\alpha\}}\Bigr]
+\frac{\omega^{3}|\widetilde{\mu}|^{2}}{3\pi\hbar\epsilon_{0}c^{3}}
\frac{\bar{n}\kappa}{k_{0}\xi},\nonumber
\end{eqnarray}
where the $\xi$ dependent terms are approximate as explained below.
\begin{figure}[h]
\includegraphics[height=2.9cm,width=7.5cm,clip]{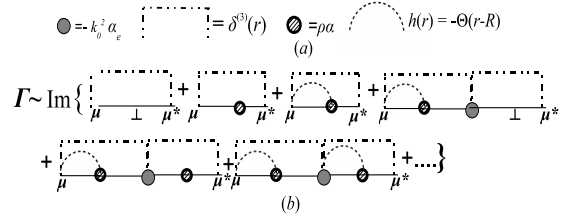}
\caption{($a$) Feynman's rules --see also Fig.\ref{fig32}($a$). ($b$) Diagrammatic representation of the total decay rate in a dilute system, Eqs.(\ref{gammatot},\ref{obapp}).}\label{fig31}
\end{figure}
Further on, invoking causality,  $\bar{n}(\omega)$ and $\kappa(\omega)$ are related through integral equations in $\omega$-modes. That is, by Kramers-Kronig relationships (KK). Therefore, one can in principle control the variations in $2\Gamma_{\perp}$ as a function of $\omega$ and so optimize its value once the variations of either $\bar{n}(\omega)$ or $l_{ext}(\omega)$ are known \cite{paperIII}.
On the other hand, in order to get full control over the total spontaneous emission we have to find similar expressions for $\Gamma_{\parallel}$. In the limit $k_{0}\xi\ll1$ we can approximate  $\chi_{\parallel}\approx\chi_{\perp}(k_{\perp})$. MG belongs indeed to this case. In such a situation it is possible to give a formula for total $\Gamma(\omega)$ in closed form once $\alpha(\omega)$ is known. As the medium surrounding the emitter is continuous beyond some distance $R=\xi$ in application of $k_{0}\xi\ll1$, Eq.(\ref{la3}) and the leading order $\xi$-dependent term of Eq.(\ref{lb1}) approximate the formulae for a real cavity in \cite{paperI} with $\chi(\omega)=\frac{\rho\alpha(\omega)}{1-\frac{1}{3}\rho\alpha(\omega)}$, yielding the second term in Eq.(\ref{lalo}) and (see also \cite{MeIII})
\begin{eqnarray}\label{laparalela}
\Gamma_{\parallel}\approx\frac{2\omega^{3}|\widetilde{\mu}|^{2}}{6\pi\hbar\epsilon_{0}c^{3}}
\Bigl\{\frac{2\bar{n}\kappa}{3}\Bigl[\frac{1}{(k_{0}\xi)^{3}}+\frac{1}{k_{0}\xi}\Bigr]
\Bigl[1+\frac{2}{(\bar{n}^{2}+\kappa^{2})^{2}}\Bigr]
\nonumber\\-\frac{\bar{n}\kappa}{k_{0}\xi}+\frac{5}{9}\Bigl[1+\frac{\kappa^{6}-\bar{n}^{6}+(\kappa^{2}-\bar{n}^{2})(4-\kappa^{2}\bar{n}^{2})}
{5(\kappa^{2}+\bar{n}^{2})^{2}}\Bigr]\Bigr\}.
\end{eqnarray}
The total decay rate is thus a function of $\omega$ and $\xi$. Optimization of $\Gamma$ with respect to $\omega,\xi$ leads this way to inhibition/enhancement of spontaneous emission.\\
 \indent On the other hand, it is
expected that not only the functional form of
$\epsilon_{e}(\omega)$ in host scatterers, but also the spatial
correlations among these, cooperate towards
inhibition/enhancement. For $\xi\sim k_{0}^{-1}$ the above
treatment fails as, in general, additional poles in Eq.(\ref{la2})
enter the problem and $\bar{n}$, $l_{ext}$ do not suffice to
compute $\Gamma$. Full knowledge of $\bar{\chi}(k)$ is needed.
Nevertheless, by invoking energy conservation, Ward's identities
for vector waves \cite{barenkov} formulated in a suitable manner
may yield integral expressions -- sum rules over $k$-modes --
relating real and imaginary parts of $\chi_{\parallel}(k)$
and $\chi_{\perp}(k)$.\\
\indent Next, we consider a simple case to study microscopically
the nature of each of the terms contributing to $\Gamma$ in
Eqs.(\ref{la1}-\ref{la3}). The setup consists of an emitter on top
of a host particle in a dilute  medium made of spherical
inclusions of polarizability $\alpha$. A similar model was studied
in \cite{Luis,LoudonI}. In this simple model, the only relevant
correlation is that given by the two-point function
$h(r)=-\Theta(r-\xi)$, $\xi$ being the radius of the exclusion
volume surrounding each scatterer. Therefore, the electrical
susceptibility reads $\chi_{\parallel,\perp}\simeq
\rho\alpha+\chi_{\parallel,\perp}^{(2)}$, where
\begin{eqnarray}
\chi^{(2)}_{\perp}(k)&=&\frac{-(\alpha\rho)^{2}k_{0}^{2}}{2}\int\textrm{d}^{3}r\:e^{i\vec{k}\cdot\vec{r}}h(r)\textrm{Tr}
\{\bar{G}^{(0)}(r)[\bar{I}-\hat{k}\otimes\hat{k}]\}\nonumber\\&=&
-\frac{(\alpha\rho)^{2}k_{0}^{2}}{2}\int\frac{\textrm{d}^{3}k'}{(2\pi)^{3}}
h(|\vec{k}'-\vec{k}|)\Bigl[G_{\perp}^{(0)}(k')\nonumber\\&+&G_{\perp}^{(0)}(k')\cos^{2}{\theta}\:+\:G_{\parallel}^{(0)}(k')\sin^{2}{\theta}\Bigr],
\label{Xiperp}
\end{eqnarray}
\begin{eqnarray}
\chi^{(2)}_{\parallel}(k)&=&-(\alpha\rho)^{2}k_{0}^{2}\int\textrm{d}^{3}r\:e^{i\vec{k}\cdot\vec{r}}h(r)\textrm{Tr}
\{\bar{G}^{(0)}(r)[\hat{k}\otimes\hat{k}]\}\nonumber\\&=&-(\alpha\rho)^{2}k_{0}^{2}\int\frac{\textrm{d}^{3}k'}{(2\pi)^{3}}
h(|\vec{k}'-\vec{k}|)\nonumber\\&\times&\Bigl[G_{\parallel}^{(0)}(k')\cos^{2}{\theta}\:+\:G_{\perp}^{(0)}(k')\sin^{2}{\theta}\Bigr],
\label{Xiparall}
\end{eqnarray}
with $\cos{\theta}\equiv \hat{k}\cdot\hat{k}'$. In the above
equations, $G^{(0)}_{\perp,\parallel}$ are the transverse
and longitudinal components of the propagator of the electric
field in free space, $\bar{G}_{\perp}^{(0)}(k)=\frac{\Delta(\hat{k})}{k_{0}^{2}-k^{2}}$,
$\bar{G}_{\parallel}^{(0)}(k)=\frac{\hat{k}\otimes\hat{k}}{k_{0}^{2}}$;
$\hat{k}$ being a unitary vector along the propagation direction
and $\Delta(\hat{k})\equiv I-\hat{k}\otimes\hat{k}$ being the
projective tensor orthogonal to the propagation direction. In
spatial space, $\bar{G}^{(0)}(r)=\frac{e^{ik_{0} r}}{-4\pi
r}\bar{I}+\bigl[\frac{1}{k_{0}^{2}}\vec{\nabla}\otimes\vec{\nabla}\bigr]\frac{e^{ik_{0}
r}}{-4\pi r}$. At leading order in $\rho\alpha$, the problem
becomes one of single scattering. Therefore, the only correlation
that matters is that of the emitter with the
host scatterers, $h(r)$.\\
\indent Pictorially, upon self-polarization cycles, $\Gamma$ is given
by the first three
 diagrams of Fig.\ref{fig31}($b$). It reads,
\begin{eqnarray}\label{gammatot}
\Gamma&=&\Gamma_{0}-\frac{2\omega^{2}|\mu|^{2}}{3\hbar\epsilon_{0}c^{2}}\Im{}\Bigl\{\textrm{Tr}\{\int\textrm{d}^{3}r_{1}
\:\bar{G}^{(0)}(\vec{r}_{1})(-k_{0}^{2}\rho\alpha)\nonumber\\&\times&
\bar{G}^{(0)}(\vec{r}_{1})\:[1-\Theta(r_{1}-\xi)]\}\Bigr\},
\end{eqnarray}
where $\Gamma_{0}\equiv\frac{\omega^{3}|\mu|^{2}}{3\pi\hbar\epsilon_{0}
c^{3}}$ is the total decay rate in vacuum disregarding self-polarization effects.
In the limit $k_{0}\xi\ll1$, we expand $\Gamma$ in powers
of $k_{0}\xi$,
\begin{eqnarray}
\Gamma\simeq\Gamma_{0}\Bigl\{1&+&\Re{\{\alpha\rho\}}[\frac{7}{6}-\frac{11}{15}(k_{0}
\xi)^{2}]\label{radapp}\\&+&\Im{\{\alpha\rho\}}[\frac{1}{(k_{0}
\xi)^{3}}+\frac{1}{k_{0}\xi}]\Bigr\}.\label{nonradpp}
\end{eqnarray}
The first two terms on the right of Eq.(\ref{radapp}) equal the
usual Lorentz-Lorenz (LL) and Onsager--B\"{o}ttcher (OB) formulae
in absence of self-polarization \cite{Lorentz,Onsager,deVries},
$\Gamma_{LL}=\Gamma_{0}\Bigl(\frac{\epsilon+2}{3}\Bigr)^{2}
\sqrt{\epsilon},\Gamma_{OB}=\Gamma_{0}\Bigl(\frac{3\epsilon}{2\epsilon+1}\Bigr)^{2}
\sqrt{\epsilon}\simeq\Gamma_{0}(1+\frac{7}{6}\Re{\{\alpha\rho\}})$,
with $\epsilon=1+\Re{\{\alpha\rho\}}$. 
A detailed analysis shows that a contribution
$\frac{\Gamma_{0}}{3}\Re{\{\alpha\rho\}}$ comes from
$\Gamma^{\parallel}$ while a contribution
$\frac{5\Gamma_{0}}{6}\Re{\{\alpha\rho\}}$ comes from
$2\Gamma^{\perp}$. We show this as follows. The vertex provided by
Eqs.(\ref{Xiperp},\ref{Xiparall}), which gives rise to coupling of
longitudinal and transverse modes  --see Fig.\ref{fig32}$(b)$,
yields a term common to $2\Gamma^{\perp}$ and
$\Gamma_{\parallel}$,
\begin{eqnarray}\label{common}
\Gamma_{0}\:\Re{\{\alpha\rho\}}k_{0}\Im{}\Bigl\{\int\frac{\textrm{d}^{3}k'}{(2\pi)^{3}}\frac{\textrm{d}^{3}k}{(2\pi)^{3}}
h(|\vec{k}'-\vec{k}|)\nonumber\\ \times
G_{\perp}^{(0)}(k')G_{\parallel}^{(0)}(k)\sin^{2}{\theta}\Bigr\}.
\end{eqnarray}
In the limit $k_{0}\xi\ll1$, the above expression reduces to
$\frac{\Gamma_{0}}{3}\Re{\{\alpha\rho\}}$. It equals the
contribution of each local field factor, $\frac{\epsilon+2}{3}$ in
$\Gamma_{LL}$ and $\frac{3\epsilon}{2\epsilon+1}$ in $\Gamma_{OB}$. On the other hand, $2\Gamma^{\perp}$
includes the term
\begin{equation}\label{raduncommon}
\Gamma_{0}\:\Re{\{\alpha\rho\}}k_{0}\Im{\Bigl\{\int\frac{\textrm{d}^{3}k}{(2\pi)^{3}}2[G_{\perp}^{(0)}(k)]^{2}\Bigr\}},
\end{equation}
which amounts to $\frac{\Gamma_{0}}{2}\Re{\{\alpha\rho\}}$.
 Its corresponding counterpart in $\Gamma_{LL}$ and $\Gamma_{OB}$ is in the bulk
factor $\sqrt{\epsilon}$. In some works only the bulk factor is associated to propagating decay rate \cite{LagvanTig,paperIII} or even to the total transverse decay rate \cite{Tip}.
Our microscopic approach reveals clear discrepancies with those interpretations.\\
\indent Moreover, one of the common terms given in Eq.(\ref{common}), $\frac{\Gamma_{0}}{3}\Re{\{\alpha\rho\}}$,
corresponds to $\Gamma_{\parallel}$ and is associated to
$\Im{\{\chi_{\parallel}/(\rho\alpha)}\}$ in Eq.(\ref{la3}). It is
therefore part of the incoherent emission which is scattered away
as bare longitudinal modes couple to bare transverse-radiation
modes at the cavity surface. The existence of the common term in
Eq.(\ref{common}) is a consequence of reciprocity. However,
while it gives rise to propagating emission when the coupling
reads $G_{\parallel}-G_{\perp}$, it
gives rise to extinction as read in opposite direction as in Fig.\ref{fig32}($b$). This longitudinal term
corresponds to the remaining local field factor in $\Gamma_{LL}$ and $\Gamma_{OB}$ at leading order. Some works \cite{tutosnow,LoudonI,Scheel} make the identification $\Gamma_{LL,OB}=2\Gamma_{LL,OB}^{\perp}$, which is erroneous according to our approach.\\
\indent In the presence of absorbtion, the electrostatic
polarizability is complex and additional non-propagating (NP)
terms in both $2\Gamma^{\perp}$ and $\Gamma^{\parallel}$
contribute to the total decay rate. These are,
\begin{eqnarray}
2\Gamma^{NP}_{\perp}&=&\Gamma_{0}\Im{\{\alpha\rho\}}k_{0}\int\frac{\textrm{d}^{3}k'}{(2\pi)^{3}}\frac{\textrm{d}^{3}k}{(2\pi)^{3}}
h(|\vec{k}'-\vec{k}|)G_{\perp}^{(0)}(k')\nonumber\\&\times&(1+\cos^{2}{\theta})G_{\perp}^{(0)}(k),\label{noradper}\\
\Gamma_{\parallel}&=&\Gamma_{0}\Im{\{\alpha\rho\}}k_{0}\int\frac{\textrm{d}^{3}k'}{(2\pi)^{3}}
\frac{\textrm{d}^{3}k}{(2\pi)^{3}}
G^{(0)}_{\parallel}(k)[h(|\vec{k}'-\vec{k}|)\nonumber\\&\times&G_{\parallel}^{(0)}(k')\cos^{2}{\theta}+\delta^{(3)}(\vec{k}-\vec{k}')
G_{\parallel}^{(0)}(k')].\label{noradparal}
\end{eqnarray}
At leading order in $k_{0}\xi$, the dominant term is that of
Eq.(\ref{noradparal}) due to absorbtion in the Coulomb-Coulomb
coupling,
$\Gamma^{NP}\approx\Gamma^{abs.}_{\parallel}\simeq\Gamma_{0}\:\Im{\{\alpha_{0}\rho\}}\frac{1}{(k_{0}
\xi)^{3}}$ \cite{Luis,LoudonI,BulloughHynne}. The remaining term $\Gamma_{0}\frac{\Im{\{\alpha\rho\}}}{k_{0}\xi}$ of Eq.(\ref{nonradpp}) contains equal contributions from $2\Gamma_{\perp}^{NP}$ and $\Gamma_{\parallel}$ in agreement with Eqs.(\ref{lalo},\ref{laparalela}).\\
\indent In absence of absorbtion, the terms in
Eqs.(\ref{noradper},\ref{noradparal}) still contribute as
considering radiative corrections in $\alpha$. These corrections
affect both the host medium and the host particle. In the former
case, radiative corrections show up as a consequence of the
coupling between transverse and longitudinal modes in single
scattering events \cite{paperI}. This is due to the 'internal'
correlation function $h_{a}(r)=\Theta(r-a)$ --see
Fig.\ref{fig32}($d$). Fig.\ref{fig32}($d$) illustrates how, for an
incoming and an outgoing photon being longitudinal, the coupling
to a transverse internal mode gives rise to radiation. Hence, the
effective vertex in Fig.\ref{fig32}($c$) with $\alpha$ including
radiative corrections.
 This kind of terms amount to scattered incoherent radiation. Eq.(\ref{noradper}) is in fact part of Eq.(\ref{lb1})
  and so of Eq.(\ref{lalo}).\\
\indent Finally, we consider the back-reaction of the host medium on the
host particle, with electrostatic polarizability $\alpha_{e}$ --eventually equal to $\alpha_{0}$.
To this aim and at first order, we have to compute the  self-polarization diagrams, i.e. $4^{th}$ to $6^{th}$ diagrams in
Fig.\ref{fig31}($b$). For the sake of simplicity, we disregard absorbtion and radiative corrections in $\alpha$ and keep leading order terms in $\alpha_{0}$ --denoted by the script $(1)$. We can approximate
\begin{equation}
\gamma^{(1)}_{\parallel}\simeq\frac{-k_{0}}{2\pi}\rho\alpha_{0}[\frac{i}{3}+\frac{1}{(k_{0}\xi)^{3}}],\:\:
2\gamma^{(1)}_{\perp}\simeq\frac{-k_{0}}{2\pi}[i+\rho\alpha_{0}\frac{5i}{6}],\label{le2}
\end{equation}
so that, in absence of self-polairzation, we obtain just $\Gamma^{(1)}_{LL}=
\Gamma^{(1)}_{OB}=\Gamma_{0}(1+\frac{7}{6}\Re{\{\alpha_{0}\rho\}})$ as given above. Plugging Eq.(\ref{le2}) into Eq.(\ref{latercera2}), we get
\begin{equation}\label{obapp}
\Gamma^{(1)}\simeq\Gamma_{LL}^{(1)}[1+\frac{4}{9}\frac{\alpha_{e}}{V_{\xi}}\rho\alpha_{0}],
\end{equation}
where $V_{\xi}$ is the volume of the cavity. Making the
identification $\epsilon=1+\alpha_{0}\rho$, one can verify the
agreement of the above expression with the OB formula at leading
order in $\alpha_{0},\alpha_{e}$ \cite{deVries,BulloughHynne},
$\Gamma_{OB}=\Gamma_{0}\Bigl(\frac{3\epsilon}{2\epsilon+1-\frac{2}{3}\frac{\alpha_{e}}{V_{\xi}}(\epsilon-1)}\Bigr)^{2}\sqrt{\epsilon}$.\\
\indent In summary, we have derived equations which relate the
spontaneous decay rate of an emitter within a virtual cavity, both
propagating --Eq.(\ref{theone})-- and non-propagating
--Eqs.(\ref{lalo},\ref{laparalela}), to the mean free path
$l_{ext}$ and the refraction index $\bar{n}$ of the host medium.
The last equation being strictly valid only in the range of validity of
MG. Because $\bar{n}$ and $l_{ext}$ depend ultimately on $\omega$
and $\xi$, it is argued that variations on $\Gamma$ towards
inhibition/enhancement can be achieved by applying KK relations
and Ward's identities. Our derivation is based on the approach of
\cite{paperI}. That is shown to include both LL and OB at leading
order in $\rho$. Microscopical analysis of the scattering events
involved in the formulae of \cite{paperI} reveals that local field
factors present longitudinal decay rate contributions. A
non-propagating decay rate has been identified. It includes both
absorbtion and incoherent scattered radiation. The latter involves
coupling of bare transverse to bare longitudinal modes which takes
place at different levels. Those are,  at the emitter cavity, at
single scatterers, at the
correlation boundary between host scatterers  and at the boundary of the sample. 
The latter was not considered in this paper.\\
%
\indent We thank S.Albaladejo, R.Carminati and J.J.Saenz for
fruitful discussions and suggestions. This work has been supported by the Spanish integrated
project Consolider-NanoLight CSD2007-00046 and the EU project
 NanoMagMa EU FP7-NMP-2007-SMALL-1.
\begin{figure}[h]
\includegraphics[height=3.4cm,width=7.2cm,clip]{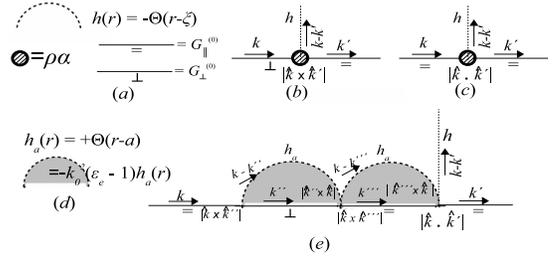}
\caption{($a$) Feynman's rules. ($b$) Common vertex derived from Eqs.(\ref{Xiperp},\ref{Xiparall}) which couples longitudinal to
transverse modes, with $|\hat{k}\times\hat{k}'|=\sin{\theta}$.
($c$) Longitudinal-longitudinal vertex from Eq.(\ref{Xiparall}), with
 $|\hat{k}\cdot\hat{k}'|=\cos{\theta}$. ($d$) Diagrammatic representation of the origin of the coupling of single scattering radiative corrections
  to longitudinal external modes. Integration of internal radiative corrections yields the vertex in Fig.($e$).}\label{fig32}
\end{figure}

\end{document}